\documentclass[aps, nofootinbib, twocolumn, showpacs, amsmath, amssymb, pra]{revtex4-1}
\usepackage{
physics,
mathtools, 
qcircuit 
}
\usepackage{graphicx} 
\newcommand{\supop}[1]{\ensuremath{\overbracket[0.1ex][0.3ex]{#1}}}

\begin{document}


\title{Constructing Smaller Pauli Twirling Sets for Arbitrary Error Channels}

\author{Zhenyu Cai}
\email{zhenyu.cai@materials.ox.ac.uk}
\author{Simon Benjamin}
\email{simon.benjamin@materials.ox.ac.uk}
\affiliation{%
 Department of Materials, University of Oxford
}%
\date{\today}

\begin{abstract}
Twirling is a technique widely used for converting arbitrary noise channels into Pauli channels in error threshold estimations of quantum error correction codes.  It is vitally useful both in real experiments and in classical quantum simulations. Minimising the size of the twirling gate set increases the efficiency of simulations and in experiments it might reduce both the number of runs required and the circuit depth (and hence the error burden). Conventional twirling uses the full set of Pauli gates as the set of twirling gates. This article provides a theoretical background for Pauli twirling and a way to construct a twirling gate set with a number of members comparable to the size of the Pauli basis of the given error channel, which is usually much smaller than the full set of Pauli gates. We also show that twirling is equivalent to stabiliser measurements with discarded measurement results, which enables us to further reduce the size of the twirling gate set.
\end{abstract}

\maketitle


\section{Introduction}
Twirling is a technique that has been long established in the quantum information literature. It was first used for mapping a diverse range of states into a canonical form in entanglement purification~\cite{bennett1996mixed, bennett1996purification}. Then it appeared again as an integral part in randomised benchmarking~\cite{knill2008randomized,magesan2011scalable} and was also used to reduce the number of experimental runs needed in quantum process tomography~\cite{emerson2007symmetrized, lu2015experimental}, both are critical in benchmarking the performance of quantum systems, especially ``Noisy Intermediate-Scale Quantum'' (NISQ) systems~\cite{preskill2018quantum}. More recently, twirling was used as means to boost the performance of NISQ through error mitigations~\cite{li2017efficient,temme2017error, endo2017practical,kandala2018extending} in which it enables a controlled increase of the gate error rates for error extrapolations. In this article, twirling is discussed as a technique for simulating noise and the impact of the noise on the performance of quantum error correction codes~\cite{geller2013efficient}.

The Gottesman-Knill theorem \cite{gottesman1998heisenberg, aaronson2004improved} states that any quantum circuits involving only Clifford gates can be perfectly simulated in polynomial time on a classical computer. One important example is the circuits used to implement quantum error correction codes. For each code, there exists an error threshold of the circuit components below which the computational error can be made arbitrarily small by scaling up the code. As we try to obtain the error thresholds of the codes, we often need to introduce various forms of noise into the circuits based on the underlying physical implementations. This noise can be viewed as extra probabilistic gates on top of the perfect Clifford gates. However, the fact that this noise can be non-Clifford means that the circuits cannot be simulated efficiently classically, i.e. numerically determining the threshold becomes intractable. 

This can be solved by twirling. Twirling means that every time we run the circuit, we conjugate the noise with an gate randomly chosen from a set of gates called the twirling set. By choosing the twirling set to be the full set of Pauli operators, we can convert any noise channel into a Pauli channel whose noise elements correspond to the Pauli basis of the original noise~\cite{dur2005standard}. Such Pauli channel approximation has been shown to be effective in error threshold estimation by Geller~\textit{et al.}~\cite{geller2013efficient} and Guti\'errez~\textit{et al.}~\cite{gutierrez2015comparison}, which justify its usage in error threshold simulation across various architectures~\cite{sarvepalli2009asymmetric, ghosh2012surface, tomita2014low, o2016silicon}.


In this article we will focus on Pauli twirling, whose twirling set is a subset of Pauli gates, with the goal of converting a given noise channel into a Pauli channel. For such a goal, twirling over the full set of Pauli operators is not always optimal. If we want to apply twirling in quantum simulations or real experiments, a twirling set with a smaller size means a lower number of simulations or experiments may be needed to get the full statistical result. Moreover, a smaller twirling set allows us to choose twirling gates that have higher fidelities and/or act on fewer qubits. This will reduce the number of errors we introduce into the system due to twirling. 

In this article, we will introduce a way to exploit the symmetries in the noise channel to reduce the size of the Pauli twirling set needed for the channel. The paper is organised as follows. In Section \ref{sect:07}, we first introduce some essential concepts for our analysis. In Section~\ref{sect:twirlinwheory}, we introduce the theory of Pauli twirling, in which we obtain the requirement on the twirling set. In Section~\ref{sect:constructW}, we show a way to construct a twirling set that satisfied the conditions that we laid out. This is followed by two examples. In Section~\ref{sect:measurement_and_twirl}, we discuss how to use stabiliser measurements to further reduce the size of our twirling set. Lastly, Section~\ref{sect:conclusion} provides a summary of our results and some possible future directions. The mathematical justification for our method of construction of the twirling set is described in the appendices, which forms an essential part of the paper.

\section{Definitions of Functions and Operations} \label{sect:07}
\subsection{The Pauli Operator Set and The $*$ Operation}\label{sect:01}
$G$ is defined to be the set of $n$-qubit Pauli operators:
\begin{align}
G = \{I,X,Y,Z\}^{\otimes n}
\end{align}

For the Pauli operator set $G$, we can define a composition rule $*$, which is the same as the usual Pauli matrix multiplication but ignoring all the $\pm 1$ and $\pm i$ factors. For one qubit we have:
\begin{align*}
X*X &= Y*Y = Z*Z = I\\
Z*Y &= Y*Z = X\\
Z*X &= X*Z = Y\\
Y*X &= X*Y = Z
\end{align*}
And any composition with the identity $I$ will just return the same operator. 

The $n$-qubit case is just the tensor product of the one-qubit case. Note that $*$ is commutative.

\subsection{Commutator Function $\zeta$}\label{sect:zeta}
For $g_i,g_j \in G$, their commutator function $\zeta(g_i,g_j)$ is defined to be:
\begin{align*}
g_ig_j = \zeta(g_i, g_j) g_jg_i
\end{align*}
i.e.
\begin{align*}
\zeta(g_i, g_j) = \begin{cases}
1\quad &\text{for }[g_i,g_j] = 0\\
-1\quad &\text{for }\{g_i,g_j\} = 0\\
\end{cases}
\end{align*}
It follows that (see Appendix~\ref{app:02})
\begin{equation}\label{eqn:03}
\begin{split}
\zeta(g_i*g_j, g_k)  = \zeta(g_ig_j, g_k) = \zeta(g_i, g_k) \zeta(g_j, g_k)\\
\zeta(g_k, g_i*g_j)  = \zeta(g_k, g_ig_j) = \zeta(g_k, g_i)\zeta(g_k, g_j)
\end{split}
\end{equation}

\section{Twirling}\label{sect:twirlinwheory}
\subsection{Super-operators and Error Channels}\label{sect:noise_channel}
We use $\supop{\quad}$ to denote a super-operator:
\begin{align*}
\left(\supop{A} + \supop{B}\right) \rho = A \rho A^\dagger + B \rho B^\dagger.
\end{align*}
A general error channel $\mathcal{E}$ is of the form:
\begin{align*}
    \mathcal{E}(\rho) = \sum_M \supop{M} \rho \quad\quad \text{with }\sum_M M^\dagger M = I .
\end{align*}
In the following sections we are going to focus on only one of the noise operators $M$.

\subsection{Exact Twirling and Random Twirling}\label{sect:twirling_background}
One can think of twirling as a super-super-operator that turns one super-operator into another. Applying exact twirling $\mathcal{T}_W$ using the twirling set $W$ on the noise operator $M$  is defined as:
\begin{align}\label{eqn:19}
\mathcal{T}_W(\supop{M})
&= \frac{1}{\abs{W}}\sum_{w \in W} \supop{w M w^\dagger}.
\end{align}
In other words, each time we run the circuit, we conjugate the noise operator $M$ with a different twirling gate $w$ from the twirling set $W$. After we iterate over the whole twirling set $W$ and take the average of the results, we effectively have process above.

The goal of twirling is to turn the noise operator $M$  into a Pauli channel:
\begin{align*}
\mathcal{T}_W(\supop{M})  = \sum_{g \in G} p_g \supop{g} .
\end{align*}
where $p_g$ is the probability of the Pauli error $g$ happening, which can be $0$.

On the other hand, in random twirling, instead of systematically iterating over the whole twirling set $W$, each run we choose a random element $w_n$ from the twirling set $W$: 
\begin{align*}
\mathcal{T}^{rand}_{W, N}(\supop{M})
&= \frac{1}{N}\sum_{n=1}^N \supop{w_n M w_n^\dagger} .
\end{align*}
At finite $N$, there will be shot noise associated with the output of random twirling due to imperfect sampling over the twirling set. The shot noise can be reduced by increasing the number of runs $N$, allowing the effect of random twirling to approach the effect of exact twirling:
\begin{align*}
\lim\limits_{N \rightarrow \infty}\mathcal{T}^{rand}_{W, N} = \mathcal{T}_W.
\end{align*}
In this paper, we will focus on exact twirling, but most of the results are also applicable to random twirling.

\subsection{One-gate Twirling}\label{sect:one-gate-twirl}
Let us consider the special case where $W = \{I, w\}$, for which $W$ only contains one extra gate other than the identity. 

We will call this a one-gate twirling operation and denote it as $\mathcal{T}_{\{I, w\}}$.

Doing nested one-gate twirling with $\mathcal{T}_{\{I, w_{1}\}}$ on top of $\mathcal{T}_{\{I, w_{2}\}}$ on top of $\mathcal{T}_{\{I, w_{3}\}}$, etc,  is equivalent to twirling with $W = \expval{w_{1}, w_{2}, \cdots}$, where $\expval{w_{1}, w_{2}, \cdots}$ denotes the full set of gates that can be generated from $\{w_{1}, w_{2}, \cdots\}$ using operation $*$.
\begin{align*}
\mathcal{T}_{\{I, w_{1}\}}\cdot\mathcal{T}_{\{I, w_{2}\}}\cdots = \mathcal{T}_{ \expval{w_{1}, w_{2}, \cdots}}
\end{align*}

\subsection{Requirements and Results of Twirling}\label{sect:02}
Now we will focus on Pauli twirling, which means our twirling set consists of only Pauli operators: $W \subseteq G$. Note that all Pauli operators are Hermitian: $w = w^\dagger$.

We can break any $n$-qubit noise operator $M$ into its Pauli basis:
\begin{align*}
M & = \frac{1}{2^n}\sum_{g\in G} \Tr(g M)g\\
& = \frac{1}{2^n}\sum_{v\in V} \Tr(v M)v
\end{align*}
where $V$ is the Pauli basis of $M$:
\begin{align*}
V = \{g \in G\ |\ \Tr(g M) \neq 0\}
\end{align*}
Substituting this into (\ref{eqn:19}) and applying it onto a state $\rho$, we have:
\begin{align}
&\mathcal{T}_W(\supop{M})\rho  = \frac{1}{\abs{W}} \frac{1}{2^{2n}} \sum_{\parbox{3.1em}{\scriptsize$v, v'$ \scriptsize$\in V$}} \Tr(v M)\Tr(v' M^\dagger) \nonumber \\ 
&\ \ \qquad\qquad\qquad \qquad\times \sum_{w\in W} w v w \rho w   v' w \label{eqn:07}
\end{align}
Now let us look at sum over $W$. Using (\ref{eqn:03}), we have
\begin{align}\label{eqn:06}
&\quad \sum_{w\in W} w v w \rho  w   v' w \nonumber\\
& = v  \rho  v'\sum_{w\in W}  \zeta(w, v)\zeta(w, v')  \nonumber\\
& = v  \rho   v'\sum_{w\in W}  \zeta(w, vv')  
\end{align}
Substituting this into (\ref{eqn:07}) we get:
\begin{widetext}
    \begin{align}
    \mathcal{T}_W(\supop{M})\rho &= \frac{1}{\abs{W}} \frac{1}{2^{2n}} \sum_{\parbox{3.1em}{\scriptsize$v, v'$ \scriptsize$\in V$}} \Tr(v M)\Tr(v' M^\dagger)v  \rho   v'\sum_{w\in W}  \zeta(w, vv')  \nonumber\\
    &= \underbrace{\frac{1}{2^{2n}} \sum_{v\in V} \abs{\Tr(v M)} ^2 \supop{v}  \rho  }_{v = v'} + \underbrace{\frac{1}{\abs{W}} \frac{1}{2^{2n}} \sum_{\parbox{3.2em}{\scriptsize$v, v'\in V$\\\scriptsize$v \neq v'$}} \Tr(v M)\Tr(v' M^\dagger) v  \rho   v'\sum_{w\in W}  \zeta(w, vv')  }_{v \neq v'}\label{eqn:01}
    \end{align}
\end{widetext}
where we have made use the fact that $\zeta(w, vv) = \zeta(w, I) = 1$.

To construct a Pauli noise channel, we want the $v\neq v'$ term to vanish (see Appendix \ref{app:necess} where we show that this is a necessary condition). This can be achieved by choosing a $W$ such that
\begin{align}\label{eqn:09}
\sum_{w\in W}  \zeta(w, vv') = 0 \quad \forall v,v' \in V \text{ and } v \neq v'
\end{align}
In such a case, the result of twirling the noise operator $M$ is just
\begin{align} \label{eqn:05}
 \mathcal{T}_W(\supop{M}) &= \frac{1}{2^{2n}} \sum_{v \in V} \abs{\Tr(v M)} ^2 \supop{v}
\end{align}
Our arguments can be easily extended to the full noise channel (Section~\ref{sect:noise_channel}) by adding $\sum_M$ before all the equations. In such case, $V$ will be re-defined as the Pauli basis needed to construct all the noise elements in the noise channel. All the other results follow.

The details of how to apply twirling on erroneous quantum components and the result of such twirling is outlined in Appendix~\ref{sect:gate_twirling}.

\section{Construction of the Twirling Set}\label{sect:constructW}

As we can see from the last section, the key to twirling is to find a twirling set $W$ that satisfy (\ref{eqn:09}) for the Pauli basis $V$ of the given noise. 

The common choice is $W = G$, the full set of Pauli operators. In such a way, for any $v \neq v'$ (i.e. $vv' \neq I$), the number of elements in $G$ that commute with $vv'$ will always equal to the number of elements that anti-commute with $vv'$, thus (\ref{eqn:09}) is always satisfied. 

Hence, if we choose $W = G$, we can transform any error channel into a Pauli channel.

However, as mentioned before, twirling with the full Pauli set is not always ideal. A systematic way to construct a smaller set of $W$ is laid out in this section, whose validity is proven in Appendix \ref{sect:twirlinggroup}, \ref{sect:constructWproof}. Note that for the steps below, compositions between elements refer to the $*$ operation defined in Section~\ref{sect:01}.

Before proceeding to the steps of construction, we need to introduce the ideas of commutator table first which is crucial to our method of construction.
\subsection{Commutator Table}\label{app:com_table}
\subsubsection{Definition}
For $A\subseteq G$, $B\subseteq G$, a commutator table $\zeta(a_i,b_j)$ is defined to be
\begin{center}
    \begin{tabular}{l |c c c}
        &$b_1$ & $b_2$ &$\cdots$\\
        \hline
        $a_1$ &$\zeta(a_1, b_1)$ &$\zeta(a_1, b_2)$ & $\cdots$ \\
        $a_2$ & $\zeta(a_2, b_1)$ &$\zeta(a_2, b_2)$&$\cdots$\\
        $\vdots$&$\vdots$& $\vdots$& $\ddots$
    \end{tabular}
\end{center}

Following (\ref{eqn:03}), we then have
\begin{equation}\label{eqn:23}
\begin{split}
\text{row composition: }\zeta(a, b_j)\zeta(a', b_j) &= \zeta(a*a', b_j)\\
\text{column composition: }\zeta(a_i, b) \zeta(a_i, b') &= \zeta(a_i, b * b')
\end{split}
\end{equation}
\subsubsection{Generator Table $\zeta(\widetilde{q}_i, \widetilde{h}_j)$}\label{sect:04}
Generator tables are just commutator tables of the form:
\begin{align*}
\zeta(\widetilde{q}_{i}, \widetilde{h}_{j}) = 1 - 2 \delta_{ij}
\end{align*}
Example generator tables for different sizes of $\widetilde{H}$ are shown in Table.\ref{table:08}.
\begin{table}[htbp]
    \centering
    {\def\arraystretch{1.3}
    \begin{tabular}{l | c }
        &$\widetilde{h}_{1}$\\
        \hline
        $\widetilde{q}_{1}$&   -1\\[1ex]
        \multicolumn{2}{c}{$\abs{\widetilde{H}} = 1$}
    \end{tabular}\qquad
    \begin{tabular}{l | c c}
        &$\widetilde{h}_{1}$&$\widetilde{h}_{2}$\\
        \hline
        $\widetilde{q}_{1}$&   -1&1 \\
        $\widetilde{q}_{2}$&   1&-1 \\[1ex]
        \multicolumn{3}{c}{$\abs{\widetilde{H}} = 2$}
    \end{tabular}\qquad
    \begin{tabular}{l | c c c}
        &$\widetilde{h}_{1}$&$\widetilde{h}_{2}$ & $\widetilde{h}_{3}$\\
        \hline
        $\widetilde{q}_{1}$&   -1 &1 & 1\\
        $\widetilde{q}_{2}$&   1 &-1& 1\\
        $\widetilde{q}_{3}$&   1& 1& -1\\[1ex]
        \multicolumn{4}{c}{$\abs{\widetilde{H}} = 3$}
    \end{tabular}\qquad $\cdots\cdots$
}
    \caption{Generator tables $\zeta(\widetilde{q}_i, \widetilde{h}_j)$ for different $\abs{\widetilde{H}}$}
    \label{table:08}
\end{table}

Note that by definition, we have
\begin{align}\label{eqn:30}
\abs{\widetilde{H}} = \abs{\widetilde{Q}}
\end{align}

The rows of a generator table cannot be obtained from composing other rows, thus the row labels $\widetilde{q}_i$ also cannot be obtained from composing other row labels. Hence, all the row labels $\widetilde{q}_i$ are independent from each other, forming a valid generating set. Similarly for the column labels $\widetilde{h}_j$, hence the name generator tables.

We can compose the columns of the generator table to obtain new columns as shown in Table~\ref{table:03}.
\begin{table}[htbp]
    \centering
    {\def\arraystretch{1.3}
    \begin{tabular}{l | c| c c c| c c c c}
        & $I$ &$\widetilde{h}_{1}$&$\widetilde{h}_{2}$ &\ $\widetilde{h}_{3}$\ &\ $\widetilde{h}_{1}*\widetilde{h}_{2}$\ &\ $\widetilde{h}_{1}*\widetilde{h}_{3}$\ &\ $\widetilde{h}_{2}*\widetilde{h}_{3}$& $\widetilde{h}_{1}*\widetilde{h}_{2}*\widetilde{h}_{3}$\\
        \hline
         $\widetilde{q}_{1}$&  1& -1 &1& 1 & -1& -1 & 1&-1\\
        $\widetilde{q}_{2}$&  1& 1& -1& 1& -1& 1 & -1&-1\\
        $\widetilde{q}_{3}$&  1& 1& 1& -1& 1 & -1 & -1&-1\\
    \end{tabular}
    }
    \caption{One of the commutator tables $\zeta(\widetilde{q}_i, h_{s, j})$ obtained by composing the columns of the generator table of size $\abs{\widetilde{H}} = 3$ in Table~\ref{table:08}. Here $ h_{s, j} \in H_S \subseteq H = \expval{\widetilde{H}}$.}
    \label{table:03}
\end{table}

\subsection{Steps to Construct $W$}\label{sect:steps}
\begin{enumerate}
    \item Decompose the noise operator $M$ to obtain its Pauli basis $V$
    \begin{align*}
    V = \{g \in G\ |\ \Tr(g M) \neq 0\}
    \end{align*}
    For a general noise channel, $V$ will be the union of the Pauli basis of all the noise elements in the noise channel.
    \item Find the following set:
    \begin{itemize}
        \item[$\widetilde{V}$: ] the generating set of $V$.
        \item[$\widetilde{V}_S$: ] the subset of elements in $\widetilde{V}$ that are used to generate elements in $V - \widetilde{V}$.
    \end{itemize}
    \item Find the smallest integer $N$ that satisfies both\footnote{The first inequality is from (\ref{eqn:21}). The second inequality is to ensure the mapping in the next step can be carried out.}
    \begin{equation*}
        \begin{split}
       N &\geq  \log_2(\abs{V}) \\
       N &\geq  \abs{\widetilde{V}_S}
        \end{split}
    \end{equation*}
    We now define a generating set $\widetilde{H}$ of size $N$ and denote the complete set that it generates as $H = \expval{\widetilde{H}}$
    \item Map elements in $V$ to elements in $H$ using the following steps:
    \begin{enumerate}
        \item Map $\widetilde{V}_S$ to a subset of elements in $\widetilde{H}$
        \item Map the elements in $V - \widetilde{V}$ to elements in $H - \widetilde{H}$ by following the composition relations of the elements in $\widetilde{V}_S$.
        \item Map elements in $\widetilde{V} - \widetilde{V}_S$ to any subset of the remaining elements in $H$ (which includes the identity) \footnote{We can do this because elements in $\widetilde{V} - \widetilde{V}_S$ are not restricted by any composition relations}.
    \end{enumerate}
    Using the steps above, we can obtain the subset of $H$ that $\widetilde{V}$ maps to, which we will denoted as $H_{\widetilde{V}}$:
    \begin{align*}
    \widetilde{v}_i \mapsto h_{\widetilde{v}, i} \quad \text{for $\widetilde{v}_i \in \widetilde{V}$ and $h_{\widetilde{v}, i} \in H_{\widetilde{V}}$}
    \end{align*}
    \item Starting with the generator table $\zeta(\widetilde{q}_i, \widetilde{h}_j)$ of size $\abs{\widetilde{H}}$, we compose its columns to get the commutator table $\zeta(\widetilde{q}_i, h_{\widetilde{v}, j})$  (See Section~\ref{sect:04}).
    \item The twirling generating set $\widetilde{W}$ is constructed such that $\zeta(\widetilde{w}_i, \widetilde{v}_j) = \zeta(\widetilde{q}_i, h_{\widetilde{v}, j})$ for all $i$ and $j$.
    \item After finding $\widetilde{W}$, we can opt to twirl the error by doing nested one-gate twirling (Section~\ref{sect:one-gate-twirl}) using the elements in $\widetilde{W}$. Or equivalently, we can twirl the error using the full set of $W = \expval{\widetilde{W}}$.
    
    Note that $\widetilde{W}$ is not unique because the generating sets are not unique.
\end{enumerate}
\subsection{An Example}\label{sect:examples}
Here we will ignore the qubit labels on the operators. e.g. $IX \equiv I_1X_2$. 
\begin{enumerate}
    \item Suppose we have noise
    \begin{align*}
    M \propto IX + IZ + YX + \frac{1}{\sqrt{2}}ZX + YY
    \end{align*}
    then the Pauli basis of $M$ is
    \begin{align*}
    V = \{IX, IZ, YX, ZX, YY\}
    \end{align*}
    \item Within $V$, the only composition relation is $YY = IZ * YX$. Hence, we have:
    \begin{align*}
    \widetilde{V}_S &= \{IZ, YX\}\\
    \widetilde{V} &= \{IX,\ IZ,\ YX,\ ZX\}
    \end{align*}
    \item The smallest integer $N$ that satisfies both
    \begin{align*}
    N &\geq \log_2(\abs{V}) = 2.58 \\
    N &\geq \abs{\widetilde{V}_S} = 2
    \end{align*}
    is $N = 3$. Hence, we will define a generating set $\widetilde{H}$ of size 3.
    \item Find the mapping $\widetilde{V} \mapsto H_{\widetilde{V}} \subseteq H = \expval{\widetilde{H}}$:
    \begin{enumerate}
        \item Map $\widetilde{V}_S$ to a subset of elements in $\widetilde{H}$:
        \begin{align*}
        \widetilde{V}_S = \{IZ, YX\} \mapsto \{\widetilde{h}_1,\widetilde{h}_2\} \subseteq \widetilde{H}
        \end{align*}
        \item Map elements in $V - \widetilde{V}$ to elements in $H - \widetilde{H}$ by following the way we use $\widetilde{V}_S$ to generate elements in $V - \widetilde{V}$:
        \begin{align*}
        V - \widetilde{V} = \{YY = IZ*YX\} \mapsto \{\widetilde{h}_1*\widetilde{h}_2\}
        \end{align*}
        \item Map $\widetilde{V} - \widetilde{V}_S$ to any subset of the remaining elements in $H$:
        \begin{align*}
        \widetilde{V} - \widetilde{V}_S = \{IX, ZX\} \mapsto \{I, \widetilde{h}_3\}
        \end{align*}
    \end{enumerate}
    Hence, we find:
    \begin{align*}
    \widetilde{V} = \{IX, IZ, YX, ZX\} \mapsto H_{\widetilde{V}} = \{I, \widetilde{h}_1, \widetilde{h}_2, \widetilde{h}_3\}
    \end{align*}
    \item Starting with the generator table of $\abs{\widetilde{H}} = 3$, we can construct the commutator table $\zeta(\widetilde{q}_i, h_{\widetilde{v}, j})$:
    \begin{center}
        {\def\arraystretch{1.3}
        \begin{tabular}{l | c| c c c}
            & $I (IX)$ &$\widetilde{h}_{1} (IZ)$&$\widetilde{h}_{2} (YX)$ &\ $\widetilde{h}_{3} (ZX)$\\
            \hline
            $\widetilde{q}_{1}$&  1& -1 &1& 1\\
            $\widetilde{q}_{2}$&  1& 1& -1& 1\\
            $\widetilde{q}_{3}$&  1& 1& 1& -1\\
        \end{tabular}
        }
    \end{center}
    In the brackets are the elements in $\widetilde{V}$ that the elements in $H_{\widetilde{V}}$ map to.
    \item Our goal is just to find $\widetilde{W}$ such that $\zeta(\widetilde{w}_i, \widetilde{v}_j) = \zeta(\widetilde{q}_i, h_{\widetilde{v}, j})$.

    A possible choice is $\widetilde{W} = \{IX, ZI, YI\}$, which produces the following commutator table 
      \begin{center}
        {\def\arraystretch{1.3}
       \begin{tabular}{l | c| c c c}
           & $IX$ &$IZ$ &$YX$ &\ $ZX$\\
           \hline
           $IX$&  1& -1 &1& 1\\
           $ZI$&  1& 1& -1& 1\\
           $YI$&  1& 1& 1& -1\\
       \end{tabular}
        }
      \end{center}
  This is the same as the commutator table in the last step.
    \item Twirling of $M$ can be achieved using nested one-gate twirling over the elements in $\widetilde{W}$:
    \begin{align*}
    \mathcal{T}_{\{I, IX\}}\cdot\mathcal{T}_{\{I, ZI\}}\cdot\mathcal{T}_{\{I, YI\}}
    \end{align*}
    Or equivalently, we can twirl over the full twirling set of
    \begin{align*}
    W &= \expval{\widetilde{W}}\\ 
    &= \{II,\ IX,\ ZI,\ YI,\ ZX,\ YX,\ XI,\ XX\}
    \end{align*}
\end{enumerate}
Using (\ref{eqn:05}), the result of twirling the noise operator $M$ is just
\begin{align*}
\mathcal{T}_W(\supop{M}) & \propto \supop{IX} + \supop{IZ} + \supop{YX} + \frac{1}{2} \supop{ZX} + \supop{YY}
\end{align*}
\subsection{Another Example with Physical Noise Operators}\label{sect:physical_example}
We will provide another example that has physical significance to illustrate the power of our method. In spin qubits, if there is a \textbf{small} fluctuating global magnetic field in the z-direction, we will have a noise operator proportional to the sum of $Z$ components of the spins. For 8 qubits we have:
\begin{align*}
M \propto \sum_{n = 1}^8 Z_n
\end{align*}
Using our methods, we have the following steps:
\begin{enumerate}
    \item The Pauli basis of $M$ is
    \begin{align*}
    V = \{Z_n |\ n \in \mathbb{N}, 1\leq n \leq 8 \}
    \end{align*}
    \item Within $V$, there are no composition relations among the elements. Hence, we have:
    \begin{align*}
    \widetilde{V}_S &= \emptyset\\
    \widetilde{V} &=  V =  \{Z_n |\ n \in \mathbb{N}, 1\leq n \leq 8\}
    \end{align*}
    \item The smallest integer $N$ that satisfies both 
    \begin{align*}
    N &\geq \log_2(\abs{V}) = 3 \\
    N &\geq \abs{\widetilde{V}_S} = 0
    \end{align*}
    is $N = 3$. Hence, we will define a generating set $\widetilde{H}$ of size 3.
    \item Using the fact that $\widetilde{V}_S = V - \widetilde{V}= \emptyset$, the following mapping $\widetilde{V} \mapsto H_{\widetilde{V}} \subseteq H = \expval{\widetilde{H}}$ can be found:
    \begin{align*}
    &\{Z_1, Z_2, Z_3, Z_4, Z_5, Z_6, Z_7, Z_8\} \mapsto \\
    &\{I,\ \widetilde{h}_1,\ \widetilde{h}_2,\ \widetilde{h}_3,\  \widetilde{h}_1* \widetilde{h}_3,\ \widetilde{h}_2*\widetilde{h}_3,\ \widetilde{h}_1*\widetilde{h}_2*\widetilde{h}_3\}
    \end{align*}
    \item Now starting with the generator table of $\abs{\widetilde{H}} = 3$, we can construct the commutator table $\zeta(\widetilde{q}_i, h_{\widetilde{v}, j })$:
    \begin{widetext}
        \begin{center}
            {\def\arraystretch{1.3}
            \begin{tabular}{l | c| c c c| c c c c}
                & $I (Z_1)$ &$\widetilde{h}_{1} (Z_2)$&$\widetilde{h}_{2}(Z_3)$ &\ $\widetilde{h}_{3}(Z_4)$\ &\ $\widetilde{h}_{1}*\widetilde{h}_{2}(Z_5)$\ &\ $\widetilde{h}_{1}*\widetilde{h}_{3}(Z_6)$\ &\ $\widetilde{h}_{2}*\widetilde{h}_{3}(Z_7)$& $\widetilde{h}_{1}*\widetilde{h}_{2}*\widetilde{h}_{3}(Z_8)$\\
                \hline
                $\widetilde{q}_{1}$&  1& -1 &1& 1 & -1& -1 & 1&-1\\
                $\widetilde{q}_{2}$&  1& 1& -1& 1& -1& 1 & -1&-1\\
                $\widetilde{q}_{3}$&  1& 1& 1& -1& 1 & -1 & -1&-1\\
            \end{tabular}
            }
        \end{center}
    \end{widetext}
    In the brackets are the elements in $\widetilde{V}$ that the elements in $H_{\widetilde{V}}$ map to.
    \item Our goal is just to find $\widetilde{W}$ such that $\zeta(\widetilde{w}_i, \widetilde{v}_j) = \zeta(\widetilde{q}_i, h_{\widetilde{v}, j})$.
    
    A possible choice is to have $\widetilde{W} = \{X_2X_5X_6X_8,\ X_3X_5X_7X_8,\ X_4X_6X_7X_8\}$, which will produce the following commutator table:
    \begin{center}
        {\def\arraystretch{1.3}
        \begin{tabular}{l | c| c c c| c c c c}
            & $Z_1$ &$Z_2$& $Z_3$ &$Z_4$ & $Z_5$ &$Z_6$& $Z_7$ &$Z_8$\\[0.7ex]
            \hline
            $X_2X_5X_6X_8$&  1& -1 &1& 1 & -1& -1 & 1&-1\\
            $X_3X_5X_7X_8$&  1& 1& -1& 1& -1& 1 & -1&-1\\
            $X_4X_6X_7X_8$&  1& 1& 1& -1& 1 & -1 & -1&-1\\
        \end{tabular}
        }
    \end{center}
    This is the same as the commutator table in the last step.
\end{enumerate}
Using (\ref{eqn:05}), the result of twirling the noise operator $M$ is just
\begin{align*}
\mathcal{T}_W(\supop{M}) & \propto \sum_{n = 1}^8 \supop{Z_n}
\end{align*}
\subsection{Expected size of $\widetilde{W}$}
Using (\ref{eqn:14}), (\ref{eqn:21}), (\ref{eqn:16}) and $\abs{W} = 2^{\abs{\widetilde{W}}}$, we have
\begin{align*}
\log_2(\abs{V}) &\leq \abs{\widetilde{W}} \leq \abs{\widetilde{V}}\\
\abs{V} &\leq  \abs{W} \leq 2^{\abs{\widetilde{V}}}
\end{align*}
Hence, unlike the full Pauli operator set whose size $4^n$ is dependent on the number of qubits $n$ that we are considering, the size of our twirling set $\abs{W}$ is only dependent on the sizes of the Pauli basis and the generating set of the Pauli basis of the particular noise channel we have. Noise arising from real physical process usually have symmetries present. Such symmetry constraints will reduce the size of the Pauli basis that builds our noise, which enable us to find a much smaller twirling set than the full Pauli set. 

One such example was shown in the last section (Section~\ref{sect:physical_example}), in which the lower bound is reached: $\abs{\widetilde{W}} = \log_2(\abs{V})$. For such noise due to the fluctuation of a global field, we have $\abs{V} = \abs{\widetilde{V}} = n$ where $n$ is the number of qubits. Hence, we have $\abs{\widetilde{W}} = \log_2(n)$. Comparing to the twirling using the full set of Pauli operators: $\abs{\widetilde{W}} = 2n$, there is an exponential reduction of the size of the twirling set.
\section{Twirling and Measurements in Stabiliser Code} \label{sect:measurement_and_twirl}
\subsection{Stabiliser code}\label{sect:stb}
In quantum error correction codes, we try to encode logical qubits into a larger number of physical qubits. All the states of the logical qubits $\ket{\psi_L}$ will live in a subspace $\mathcal{V_S}$ of the full quantum space of the physical qubits. We will call $\mathcal{V_S}$  the code subspace. Quantum states that live outside the code subspace can be detected as erroneous states and might be corrected by projecting (or transforming) back to the code subspace. 

If we have a given code subspace $\mathcal{V_S}$. Then the stabiliser set $S \subseteq G$ is defined to be:
\begin{align*}
S = \{s \in G\ |\ s \ket{\psi_L} = \ket{\psi_L}\quad \forall \ket{\psi_L}\in \mathcal{V_S}\}
\end{align*}
Hence, for any $s \in S$ we have
\begin{align*}
\left(\frac{1 + s}{2}\right) \ket{\psi_L} &= \ket{\psi_L}\\
\left(\frac{1 - s}{2}\right) \ket{\psi_L} &= 0 
\end{align*}
\subsection{Equivalence of one-gate twirling and stabiliser checks}
\subsubsection{one-gate twirling}
For a given noise operator $M$ and a given one-gate twirling set $W = \{I, w\}$, we can write
\begin{align*}
M = M_+ + M_-
\end{align*}
where $M_+$ contains all the Pauli basis elements in $M$ that commute with $w$, $M_-$ contains all the Pauli basis elements in $M$ that anti-commute with $w$. 

Then using (\ref{eqn:19}), we have:
\begin{align}
\mathcal{T}_{\{I, w\}}(\supop{M})\rho & = \frac{1}{2} \left[\supop{M} \rho + \supop{\left(wMw\right)} \rho\right]\nonumber\\
& = \frac{1}{2} \left[\supop{\left(M_+ + M_-\right)} \rho + \supop{\left(M_+ - M_-\right)} \rho\right]\nonumber\\
& = \frac{1}{2} \left[M_+ \rho M_+^\dagger +  M_- \rho M_-^\dagger\right]\label{eqn:29}
\end{align}
i.e. an one-gate twirl $W = \{I, w\}$ will decohere between the components in $M$ that commute with $w$ and the components that anti-commute with $w$.
\subsubsection{Stabiliser Checks}
For a given noise operator $M$ and a given stabiliser $s$, we can write
\begin{align*}
M = M_+ + M_-
\end{align*}
where $M_+$ contains all the Pauli basis elements in $M$ that commute with $s$, $M_-$ contains all the Pauli basis elements in $M$ that anti-commute with $s$. 

Then if such noise happens on $\ket{\psi_L}$, and we do an $s$ stabiliser check on it, then we have
\begin{align*}
s M \ket{\psi_L}& = \underbrace{\frac{1 + s}{2} \left(M_+ + M_-\right)\ket{\psi_L}}_{\text{projection onto $+1$ state}} + \underbrace{\frac{1 - s}{2} \left(M_+ + M_-\right)\ket{\psi_L}}_{\text{projection onto $-1$ state}}\\
& = \left[M_+\frac{1 + s}{2} \ket{\psi_L} + M_-\frac{1 - s}{2} \ket{\psi_L}\right] \\
&\quad + \left[M_+\frac{1 - s}{2} \ket{\psi_L} + M_-\frac{1 + s}{2} \ket{\psi_L}\right]\\
& = \underbrace{M_+\ket{\psi_L}}_{\parbox{5em}{\scriptsize projection onto $+1$ state}} + \underbrace{M_-\ket{\psi_L}}_{\parbox{5em}{\scriptsize projection onto $-1$ state}}
\end{align*}
Here we can see that an $s$ stabiliser check that gives $+1$ result will collapse the state into $M_+ \ket{\psi_L}$ (up to a normalising constant), an $s$ stabiliser check that gives $-1$ result will collapse the state into $M_- \ket{\psi_L}$ (up to a normalising constant).

If we discard the information about result of the $s$ stabiliser check, then our error channel after the $s$ stabiliser check becomes:
\begin{align}\label{eqn:18}
\mathcal{S}_{s}(\supop{M})\rho &= \frac{1}{2} \left[M_+\rho M_+^\dagger + M_-\rho_L M_-^\dagger\right]
\end{align}
where $\rho = \ket{\psi_L}\bra{\psi_L}$.

We can follow similar analysis even if there is a Pauli error $g$ on the logical state $\ket{\psi_L} \rightarrow g \ket{\psi_L}$. The extra Pauli error may swap the $\pm 1$ stabiliser check outcome, but will not change our error channel in (\ref{eqn:18}).

Comparing (\ref{eqn:18}) to (\ref{eqn:29}), we have:
\begin{align*}
\mathcal{T}_{\{I, s\}} \equiv \mathcal{S}_{s}
\end{align*}
Hence, when we have a error $M$ occurring on top of $g \ket{\psi_L}$, twirling with $W = \{I, s\}$ is equivalent to performing a $s$ stabiliser check and throwing away the result.
\subsection{Combining stabiliser check with twirling}
As mentioned in the last step of Section~\ref{sect:steps}, a given noise operator $M$ can be twirled by doing nested one-gate twirling using the elements in the twirling generating set $\widetilde{W}$. In the last section, we have shown that the $s$-base stabiliser measurement is equivalent to the one-gate twirling with $W = \{I, s\}$. Hence, we can use the $s$ stabiliser check as a substitute for element $s$ in the $\widetilde{W}$ to further reduce the size of $\widetilde{W}$.

This is best shown through a simple example.

Suppose we have the following circuit:
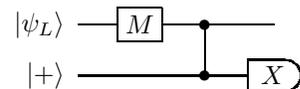
\begin{align*}
\Qcircuit @C=1.5em @R=.7em {
    &\lstick{\ket{\psi_L}} & \gate{M} &\ctrl{1}&\qw\\
    &\lstick{\ket{+}}      & \qw      &\ctrl{0}   &\measureD{X}
} 
\end{align*}
Here $\ket{\psi_L}$ is stabilised by $Z$, i.e. $\ket{\psi_L} = \ket{0}$ \footnote{Here we have one physical qubit and one stabiliser, hence the state of the physical qubit is fixed by the stabiliser measurement, i.e. we encode 0 logical qubits.}. In this circuit, we are effectively doing a $Z$ stabiliser measurement on $\ket{\psi_L}$, with a noise
\begin{align*}
M \propto I + X + Y + Z
\end{align*}
occurring in the circuit. 

Now if we go through Section~\ref{sect:steps}, we will obtain the twirling generating set $\widetilde{W}$ needed to twirl $M$ as follows:
\begin{align*}
\widetilde{W} = \{X, Z\}
\end{align*}
which means we need the following nested one-gate twirling circuit to turn $M$ into Pauli errors:
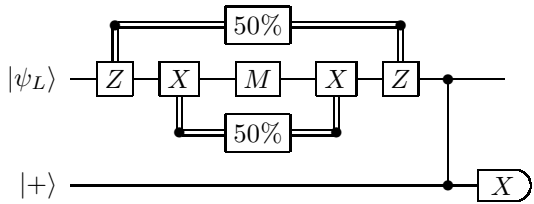
\begin{align*}
\Qcircuit @C=1em @R=.7em {
    && \control{0} & \cw& *+[F]{50 \%}\cw & \cw &\control{0}\cw &&\\
    &\lstick{\ket{\psi_L}}& \gate{Z}\cwx{-1} & \gate{X} & \gate{M} &\gate{X} & \gate{Z}\cwx{-1} &\ctrl{2}&\qw\\
    &&& \control{0}\cwx{-1} &*+[F]{50 \%}\cw &\control{0}\cwx{-1}\cw & &&\\
    &\lstick{\ket{+}}     & \qw      & \qw      & \qw      & \qw     & \qw      &\ctrl{0}   &\measureD{X}
} 
\end{align*}
where the pair of $X$ will be applied with $50\%$ probability, similarly and independently for the pair of $Z$.

However, if we discard the information specifying the result of the $Z$ stabiliser check, then as argued before, the $Z$ stabiliser check will have the same effect as the $Z$-twirling. Hence, we can turn $M$ into Pauli errors with just the $X$-twirling:
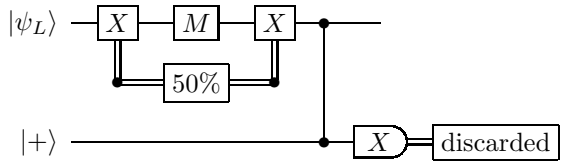
\begin{align*}
\Qcircuit @C=1em @R=.9em {
    &\lstick{\ket{\psi_L}}&  \gate{X} & \gate{M} &\gate{X}  &\ctrl{2}&\qw\\
    && \control{0}\cwx{-1} & *+[F]{50 \%}\cw &\control{0}\cwx{-1}\cw  &&\\
    &\lstick{\ket{+}}          & \qw      & \qw      & \qw     &\ctrl{0}   &\measureD{X}&*+[F]{\text{discarded}}\cw
} 
\end{align*}
This example shows how stabiliser measurements with thrown-away results can lead to a smaller set of twirling gates.

\section{Conclusion and future works}\label{sect:conclusion}
In this paper, we found the necessary and sufficient conditions for a set of twirling gate to turn a given noise operator into a Pauli channel form. We then demonstrated a way to construct the smallest twirling set that satisfies the conditions. The size of the twirling set we obtained is lower-bounded by the size of the Pauli basis of the noise operator, and upper bounded by $2^{\abs{\widetilde{V}}}$ where $\widetilde{V}$ is the generating set of the Pauli-basis of the noise operator. We showed that there can be an exponential reduction in the number of twirling gates in some cases. Our arguments can be easily extended to a general noise channel. In addition, we showed that in the case of stabiliser codes, we can replace elements in the generating set of the twirling set with existing stabiliser measurements to further reduce the size of the twirling set. 

For twirling of a given noise operator, we have not proven the twirling set we obtained is the smallest possible. Hence, any further investigations can look into such a proof or even constructing a smaller twirling set than ours.

For a general noise channel, the simple generalisation mentioned in Section~\ref{sect:steps} can indeed produce a twirling set smaller than the full set of Pauli operators. However, it is not the smallest possible set since we have not made use of the fact that different noise elements are inherently separated. To obtain the optimal twirling set, we need to study the following property of twirling: if we know a twirling set that can twirl the noise operator $M$, and we know another twirling set that can twirl the noise operator $N$, then what is the twirling set that can twirl the noise channel $\supop{M} + \supop{N}$? Similarly, we can also ask what is the twirling set that can twirl the noise operator $MN$, which is essential in finding a single twirling operation that can twirl several consecutive erroneous components. We hope that this article will provide a framework for further explorations of properties of twirling like the two mentioned above.

In this paper, we have only focused on using Pauli twirling to convert error channels into Pauli channels for error threshold estimation. There is also Clifford twirling, which converts error channels into depolarising channels instead. Clifford twirling can be viewed as symplectic twirling on top of Pauli twirling~\cite{dehaene2003clifford}, so we can easily apply our arguments to the Pauli twirling step. Clifford twirling is integral to Clifford randomised benchmarking~\cite{knill2008randomized,magesan2011scalable} and is also used in quantum process tomography to reduce the number of experiments that we need to run exponentially~\cite{emerson2007symmetrized, lu2015experimental}. We cannot apply our techniques directly to both of these areas since we do not know the form of the quantum process that we want to twirl. However, our analysis might provide a basis for finding a reduced twirling set for the case in which some characteristics of the quantum process are known, but not the full model.

\section*{Acknowledgements}
The authors thank Y. Li for valuable discussions. ZC acknowledges support from Quantum Motion Technologies Ltd. SCB acknowledges support from ESPRC grant EP/M013243/1 (the NQIT Quantum Hub).

\newpage
\appendix
\section{Twirling of Gate Noise}\label{sect:gate_twirling}
In real circuits, a noise operator is not a physical gate, hence it is impossible to bracket the noise operator with twirling gates. We can instead bracket the source of the noise with twirling gates..

Suppose we want to apply a gate $C$, but an error $M$ occurs after gate $C$ with a probability $p$:
\begin{align}\label{eqn:04}
\mathcal{C}_{e}(\rho) =   (1-p)\supop{C}\rho + p \supop{MC} \rho.
\end{align}
Now for the noisy part of the process $\supop{MC} \rho$, we want to twirl the noise $M$.
\begin{align}
\mathcal{C}_{e}(\rho) & \xrightarrow{twirling} (1-p)\supop{C}\rho + p  \mathcal{T}(\supop{M})\supop{C}\rho \label{eqn:08}\\
& = (1-p)\supop{C}\rho + p  \frac{1}{\abs{W}}\sum_{w \in W} \supop{\left(w M w\right) C} \rho\nonumber \\
& =  (1-p)\supop{C}\rho + p  \frac{1}{\abs{W}}\sum_{w \in W} \supop{w MC\left(C^\dagger wC\right)} \rho \nonumber
\end{align}
which is just the following circuit:
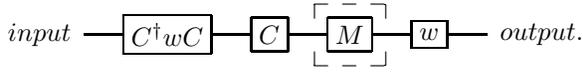
\begin{align*}
\Qcircuit @C=1.5em @R=.7em {
    &\lstick{input} &\gate{C^\dagger w C}&\gate{C}& \gate{M}&\gate{w}&\rstick{output.}\qw
    \gategroup{1}{5}{1}{5}{1em}{--}
} 
\end{align*}
Hence, by bracketing the erroneous gate $C$ with the twirling gate $w$ and its complementary gate $C^\dagger w C$, we are effectively twirling the noise $M$ coming out of $C$.

Substituting (\ref{eqn:05}) into (\ref{eqn:08}) we have
\begin{align*}
\mathcal{C}_{e}(\rho)  
\xrightarrow{twirling}  (1-p)\supop{C}\rho + \frac{p}{2^{2n}} \sum_{v \in V} \abs{\Tr(v M)}^2 \supop{v C} \rho.
\end{align*}
Hence, after twirling, $\mathcal{C}_{e}(\rho)$ becomes a error channel with Pauli error $v \in V$ happening with the probability $\frac{p \abs{\Tr(v M)}^2}{2^{2n}}$.

\section{Property of $\zeta$}\label{app:02}
For $g,g' \in G$, $\zeta(g,g')$ is defined to be:
\begin{align*}
gg' = \zeta(g, g') g'g
\end{align*}
i.e.
\begin{align*}
\zeta(g, g') = \begin{cases}
1\quad &\text{for }[g,g'] = 0\\
-1\quad &\text{for }\{g,g'\} = 0\\
\end{cases}
\end{align*}
We then have the following properties:
\begin{itemize}
    \item $\zeta(g, g')^{-1} = \zeta(g, g')$\\
    
    \textbf{Proof:}
    
    Since $\zeta(g, g') = \pm 1$.
    \\
    \item $\zeta(g, g') = \zeta(g', g)$\\
     
    \textbf{Proof:}
    \begin{align*}
    gg' &= \zeta(g, g') g'g\\
    \zeta(g, g')^{-1}gg' &= g'g
    \end{align*}
    Hence, $\zeta(g, g')^{-1} = \zeta(g', g)$. Then using $\zeta(g, g')^{-1} = \zeta(g, g')$, we have $ \zeta(g', g) =  \zeta(g, g')$.
    \\
    \item $\zeta(g, g_1g_2) = \zeta(g, g_1)\zeta(g, g_2)$\\
     
    \textbf{Proof:}
    \begin{align*}
    gg_1g_2 &= \zeta(g, g_1g_2) g_1g_2g\\
    \zeta(g, g_1)g_1gg_2 &=  \zeta(g, g_1g_2) g_1g_2g\\
    \zeta(g, g_1)\zeta(g, g_2)g_1g_2g &= \zeta(g, g_1g_2) g_1g_2g\\
    \zeta(g, g_1)\zeta(g, g_2) &= \zeta(g, g_1g_2)
    \end{align*}
    \\
    \item $\zeta(g, cg') = \zeta(g, g')$ for any complex number $c$.\\
     
    \textbf{Proof:}
    \begin{align*}
    gcg' &= \zeta(g, cg') cg'g\\
    gg' &= \zeta(g, cg') g'g\\
    \zeta(g, cg') & = \zeta(g, g')
    \end{align*}
    From this it immediately follows that:
    \begin{align*}
    \zeta(g * g', g'') &= \zeta(cgg', g'')\\
    &= \zeta(gg', g'')
    \end{align*}
    where $*$ is the operation we defined in Section.\ref{sect:01}.
\end{itemize}

\section{Proof that (\ref{eqn:09}) is a necessary condition}\label{app:necess}
(\ref{eqn:01}) can be written as
\begin{align*}
\mathcal{T}(\supop{M})\rho  &=  \sum_{g, g'\in V} \alpha_{gg'}  g \rho  g'
\end{align*}
If we indeed can transform this into a Pauli channel, then we have:
\begin{align*}
\sum_{g, g'\in G} \alpha_{gg'}  g \rho  g' &= \sum_{g''\in G} \beta_{g''}  g'' \rho g''
\end{align*}
\begin{align*}
\sum_{g} \left( \alpha_{gg} - \beta_{g} \right) g \rho g + \sum_{g \neq g', g, g'\in G} \alpha_{gg'}  g \rho g' &= 0
\end{align*}

This should be valid for any $\rho$. Using the Choi-Jamiolkowski isomorphism on the process, we have:
\begin{align*}
\sum_{g} \left(\alpha_{gg} - \beta_{g} \right) \ket{g} \bra{g} + \sum_{g \neq g', g, g'\in G} \alpha_{gg'}  \ket{g} \bra{g'} &= 0
\end{align*}
In terms of Pauli basis states $\{\ket{g}\}$, the LHS is just a matrix with $\left(\alpha_{gg} - \beta_{g} \right)$ at the diagonal and $\alpha_{vv'}$ at the off diagonal. Hence, the only way for the equation to be valid is when:
\begin{align*}
\beta_{g} &= \alpha_{gg} \\
\alpha_{gg'} &= 0
\end{align*}
which is equivalent to (\ref{eqn:09}).

\section{Properties of Commutator Tables}\label{sect:twirlinggroup}
This section proves some properties of commutator tables (See Section~\ref{app:com_table}), which is crucial in proving our construction of twirling set is valid in Appendix \ref{sect:constructWproof}.

Note that in this section, whenever we talk about composition, we are referring to the operation $*$ (Section \ref{sect:01}).

\subsection{Row Commutator Group}\label{sect:Q}
\subsubsection{Homomorphic mapping}
For any $A \subseteq G$, we can construct a commutator table $\zeta(g_i, a_j)$ with $g_i \in G$. Its rows form a set $R_A$:
\begin{align*}
R_A = \{\zeta(g, a_j)\ |\ g\in G\}
\end{align*}
From (\ref{eqn:23}), we have:
\begin{align}\label{eqn:24}
\zeta(g * g', a_j) = \zeta(g, a_j) \zeta(g', a_j)\quad \forall g,g' \in G 
\end{align}
Note that The Pauli operator set $G$ is a group under the composition rule $*$. (\ref{eqn:24}) means that there exists a homomorphic mapping: $G \mapsto R_A$. 

Hence, \textbf{$R_A$ is a also a group}.

\subsubsection{Quotient sets}\label{sect:quotient_set}
The kernel subgroup of the homomorphic mapping $G \mapsto R_A$ by definition is
\begin{align*}
K_A = \{g\in G \ |\ \zeta(g, a_j)  = (1, 1, 1, \cdots) = \vec{1}\}
\end{align*}
i.e. it maps to the set of rows in the commutator table $\zeta(g_i, a_j)$ that only contains $1$.

$K_A$ will partition $G$ into $\frac{\abs{G}}{\abs{K_A}}$ cosets. For $g \in G$ that are within the same coset $gK_A$, their corresponding row vector $\zeta(g, a_j)$ will have the same value.

We will define the \textbf{quotient set} (not group) $Q_A$ as a set that has one and only one element from each coset of $K_A$. Hence, the set of row vectors $ \{\zeta(q_a, a_j)\ |\ q_a\in Q_A\}$ will contain one and only one element for each possible row vector value.

\subsubsection{Sum of rows vectors that maps to the quotient set}
For a given $a \neq I$, the number of elements in $G$ that commute with $a$ will always equal to the number of elements that anti-commute with $a$. Hence, we have
\begin{align*}
\sum_{g\in G}  \zeta(g, a) &= 0 \quad \forall a \in A \text{ and } a\neq I
\end{align*}
As we mentioned before, all $g \in G$ that are within the same coset will have the same row vector value $\zeta(g, a_j)$. Hence, the sum over $\sum_{g\in G}$ can be divided into the sum within the same coset, which is a sum over $\abs{K_A}$ identical row vectors, and the sum over all different cosets. Hence, for all $a\neq I$, we have
\begin{align*}
\sum_{g\in G} \zeta(g, a) &= 0 \\
\abs{K_A}\sum_{q_a\in Q_A} \zeta(q_a, a) &= 0\\
\sum_{q_a\in Q_A} \zeta(q_a, a) &= 0
\end{align*}
i.e.
\begin{align}
\sum_{q_a\in Q_A} \zeta(q_a, a) = 0\quad \forall\ a \in A \text{ and } a\neq I \label{eqn:17}
\end{align}

\subsection{Quotient Table $\zeta(q_i,h_j)$}\label{sect:Qtable}
\subsubsection{Composing the rows of the generator tables}
The definition of a generator table $\zeta(\widetilde{q}_i, \widetilde{h}_j)$ is laid out in Section~\ref{sect:04}.

We will define $Q$ to be the full set of elements that can be generated from $\widetilde{Q}$:
\begin{align*}
Q = \expval{\widetilde{Q}}
\end{align*}
Just like how we can generate $q \in Q$ using $\widetilde{q} \in \widetilde{Q}$,  we can generate new rows using (\ref{eqn:23}):
\begin{align*}
q = \widetilde{q}*\widetilde{q}' \mapsto \zeta(q, \widetilde{h}_j) = \zeta(\widetilde{q}*\widetilde{q}', \widetilde{h}_j) = \zeta(\widetilde{q}, \widetilde{h}_j)\zeta(\widetilde{q}', \widetilde{h}_j)
\end{align*}
By composing rows in Table.\ref{table:08} in every possible way, we obtain the new commutator tables $\zeta(q_i, \widetilde{h}_j)$ for different $\abs{\widetilde{H}}$ in Table.\ref{table:07}.
\begin{table}[htbp]
    \centering
    {\def\arraystretch{1.3}
    \begin{tabular}{l | c }
        &$\widetilde{h}_{1}$\\
        \hline
        $q_{1}$&   1\\
        $q_{2}$&   -1\\[1ex]
        \multicolumn{2}{c}{$\abs{\widetilde{H}} = 1$}
    \end{tabular}\qquad
    \begin{tabular}{l | c c}
        &$\widetilde{h}_{1}$&$\widetilde{h}_{2}$\\
        \hline
        $q_{1}$&   1&1 \\
        $q_{2}$&   1&-1 \\
        $q_{3}$&   -1&1 \\
        $q_{4}$&   -1&-1\\[1ex]
        \multicolumn{3}{c}{$\abs{\widetilde{H}} = 2$}
    \end{tabular}\qquad
    \begin{tabular}{l | c c c}
        &$\widetilde{h}_{1}$&$\widetilde{h}_{2}$ & $\widetilde{h}_{3}$\\
        \hline
        $q_{1}$&   1 &1 & 1 \\
        $q_{2}$&   1 &1& -1\\
        $q_{3}$&   1& -1& 1\\
        $q_{4}$&   1 &-1& -1\\
        $q_{5}$&   -1 &1 & 1\\
        $q_{6}$&   -1 &1& -1 \\
        $q_{7}$&   -1& -1& 1\\
        $q_{8}$&   -1& -1& -1\\[1ex]
        \multicolumn{4}{c}{$\abs{\widetilde{H}} = 3$}
    \end{tabular}\qquad $\cdots\cdots$
    }
    \caption{Commutator table $\zeta(q_i, \widetilde{h}_j)$ for different $\abs{\widetilde{H}}$}
    \label{table:07}
\end{table}

We can see the rows of $\zeta(q_i, \widetilde{h}_j)$ consist of all possible values of $(\pm 1, \pm 1 \cdots )$ vectors of length $\abs{\widetilde{H}}$. Looking back at the definition of quotient sets in Section.\ref{sect:quotient_set},  we realise that \textbf{$Q$ is just the quotient set of $\widetilde{H}$}.

\subsubsection{Composing the columns of $\zeta(q_i, \widetilde{h}_j)$}
We will define  $H$ to be the full set of elements that can be generated from $\widetilde{H}$:
\begin{align}\label{eqn:20}
H = \expval{\widetilde{H}}
\end{align}
Just like how we can generate $h \in H$ using $\widetilde{h} \in \widetilde{H}$,  we can generate new columns using (\ref{eqn:23}):
\begin{align*}
    h = \widetilde{h}*\widetilde{h}' \mapsto \zeta(q_i, h) = \zeta(q_i, \widetilde{h}*\widetilde{h}') = \zeta(q_i, \widetilde{h}) \zeta(q_i, \widetilde{h}')
\end{align*}
If the row $\zeta(g, \widetilde{h}_j) = (1,1,\cdots, 1)$, then the row $\zeta(g, h_j) = (1,1,\cdots,1)$. Hence, $H$ and $\widetilde{H}$ share the same kernel subgroup (see Section.\ref{sect:Q}), and thus the same quotient set $Q$.

By composing all possible columns of the commutator tables in Table.\ref{table:07}, we obtain Table.\ref{table:02}.
\begin{table}[htbp]
    \centering
    {\def\arraystretch{1.3}
    \begin{tabular}{l| c | c }
        &$I$&$\widetilde{h}_{1}$\\
        \hline
        $q_{1}$&  1& 1\\
        $q_{2}$&  1& -1\\[1ex]
        \multicolumn{2}{c}{$\abs{\widetilde{H}} = 1$}
    \end{tabular}\qquad
    \begin{tabular}{l | c | c c | c}
        &$I$&$\widetilde{h}_{1}$&$\widetilde{h}_{2}$& $\widetilde{h}_{1}*\widetilde{h}_{2}$\\
        \hline
        $q_{1}$&  1& 1&1 & 1\\
        $q_{2}$&  1& 1&-1 & -1\\
        $q_{3}$&  1& -1&1 & -1\\
        $q_{4}$&   1&-1&-1& 1\\[1ex]
        \multicolumn{4}{c}{$\abs{\widetilde{H}} = 2$}
    \end{tabular}\\
    
    \begin{tabular}{l | c| c c c| c c c c}
        & $I$ &$\widetilde{h}_{1}$&$\widetilde{h}_{2}$ &\ $\widetilde{h}_{3}$\ &\ $\widetilde{h}_{1}*\widetilde{h}_{2}$\ &\ $\widetilde{h}_{1}*\widetilde{h}_{3}$\ &\ $\widetilde{h}_{2}*\widetilde{h}_{3}$\ &\  $\widetilde{h}_{1}*\widetilde{h}_{2}*\widetilde{h}_{3}$\\
        \hline
        $q_{1}$&  1& 1 &1 & 1 & 1 &1 & 1 &1\\
        $q_{2}$&  1& 1& 1& -1& 1 & -1 & -1&-1\\
        $q_{3}$&  1& 1& -1& 1& -1& 1 & -1&-1\\
        $q_{4}$&  1& 1 &-1& -1& -1 & -1 & 1&1\\
        $q_{5}$&  1& -1 &1& 1 & -1& -1 & 1&-1\\
        $q_{6}$&  1& -1& 1& -1& -1& 1 & -1&1\\
        $q_{7}$&  1& -1&-1& 1 & 1 & -1& -1&1\\
        $q_{8}$&  1& -1 &-1 & -1& 1 &1&1&-1\\[1ex]
        \multicolumn{7}{c}{$\abs{\widetilde{H}} = 3$}
    \end{tabular}
    }
    \caption{Quotient table $\zeta(q_i, h_j)$ for different $\abs{\widetilde{H}}$}
    \label{table:02}
\end{table}

Due to the symmetry between $\widetilde{Q}$ and $\widetilde{H}$, we know that $H$ is also the quotient set of $\widetilde{Q}$. Hence, $\zeta(q_i, h_j)$ is called a quotient table. 

\subsubsection{Shape of quotient tables}\label{sect:shapeQtable}
Any $h \in H$ that can be generated from $ \widetilde{h}_{ i} \in \widetilde{H}$ can be written as
\begin{align*}
h = \prod_{i = 1}^{\abs{\widetilde{H}}} \widetilde{h}_{ i}^{\alpha_i}
\end{align*}
where the $\prod$ and the exponentials here are all defined in terms of operation $*$.

$ \widetilde{h}_{ i}*\widetilde{h}_{i} = I$ for all $\widetilde{h}_{ i} \in \widetilde{H} \subseteq G$, thus $\alpha_i$ is either $0$ or $1$. Hence, there are $2^{\abs{\widetilde{H}}}$ possible choice for $\{\alpha_i\}$, which means 
\begin{align*}
\abs{H} = 2^{\abs{\widetilde{H}}}
\end{align*}
Similarly
\begin{align*}
\abs{Q} = 2^{\abs{\widetilde{Q}}} = 2^{\abs{\widetilde{H}}}
\end{align*}
where we have used (\ref{eqn:30}).

Hence, for a quotient table
\begin{equation}\label{eqn:22}
\begin{split}
\text{No. of rows} & = \abs{Q} = 2^{\abs{\widetilde{H}}}  \\
\text{No. of columns}& = \abs{H} = 2^{\abs{\widetilde{H}}} 
\end{split}
\end{equation}
\subsubsection{Sum of rows in a quotient table}
From (\ref{eqn:17}), with $A = H$, we have
\begin{align}
\sum_{q\in Q} \zeta(q, h) = 0\quad \forall\ h \in H \text{ and } h\neq I \label{eqn:10}
\end{align}
When we compose any two different elements in $H$, we will just get another non-identity elements in $H$:
\begin{align*}
h_{i} = h_{j} * h_{k} \qquad \text{where} \ h_{i}, h_{j}, h_{k} \in H,\  h_{j}\neq h_{k},\ h_{i} \neq I
\end{align*}
Substituting into (\ref{eqn:10}), we have
\begin{align}\label{eqn:13}
\sum_{q\in Q} \zeta(q, h_{j} * h_{k}) = 0\quad \forall\ h_{j},\ h_{k} \in H \text{ and } h_{j}\neq h_{k}
\end{align}

\section{Constructing the Twirling Set $W$}\label{sect:constructWproof}
\subsection{Another View on Requirements of Twirling Set}\label{Ttable}
A noise channel can be decomposed into its Pauli basis $V$. 

In this notation, to fully twirl the noise, we need the twirling set $W$ to satisfy the following equation (using (\ref{eqn:09}) and (\ref{eqn:03})):
\begin{align}
\sum_{w\in W}  \zeta(w, v * v') &= 0 \quad \forall v, v' \in V \text{ and } v \neq v'\label{eqn:02}
\end{align}
which is just (\ref{eqn:13}) with the following \textbf{bijective} mappings:
\begin{align*}
W &\mapsto Q\\
V &\mapsto H_V \subseteq H
\end{align*}
which can be simplified to
\begin{align*}
\widetilde{W} &\mapsto Q_{\widetilde{W}} \subseteq Q\\
\widetilde{V} &\mapsto H_{\widetilde{V}} \subseteq H.
\end{align*}
Remember that $Q = \expval{\widetilde{Q}}$. If we want to find the \textbf{smallest} $W$ that maps to $Q$, the only way is to have $Q_{\widetilde{W}} = \widetilde{Q}$ and $W = \expval{\widetilde{W}}$. Hence, our requirement on the twirling set becomes finding the following mappings
\begin{equation}\label{eqn:15}
\begin{split}
\widetilde{W} &\mapsto \widetilde{Q}\\
\widetilde{V} &\mapsto H_{\widetilde{V}} \subseteq H.
\end{split}
\end{equation}
The way to find such mapping is outlined in Section~\ref{sect:steps}.

Looking back at the definition of one-gate twirling in Section \ref{sect:one-gate-twirl}, we realise that doing twirling with $W$ is equivalent to doing nested one-gate twirling with the elements in $\widetilde{W}$:
\begin{align*}
\prod_{\widetilde{w} \in \widetilde{W}} \mathcal{T}_{\{I, \widetilde{w}\}} = \mathcal{T}_{W}.
\end{align*}
Hence, after find the $\widetilde{W}$ that satisfy the mapping in (\ref{eqn:15}), we can fully twirl the given error $M$ by doing nested one-gate twirling with all the elements in $\widetilde{W}$.

\subsection{Size Matching}
To achieve the mapping in (\ref{eqn:15}), we need to ensure the sizes of the sets match each other:
\begin{equation}\label{eqn:14}
\begin{split}
\abs{\widetilde{W}} &= \abs{\widetilde{Q}}= \abs{\widetilde{H}}\\
\abs{V} &=\abs{H_V} \leq \abs{H} = 2^{\abs{\widetilde{H}}} 
\end{split}
\end{equation}
where we have used (\ref{eqn:30}) and (\ref{eqn:22}).

From the first equation, we know that minimising $\abs{\widetilde{W}}$ is the same as minimising $\abs{\widetilde{H}}$. 

From the second equation, we know that
\begin{align}
\abs{V} &\leq 2^{\abs{\widetilde{H}}} \label{eqn:21}
\end{align}
Since we are looking for the \textbf{smallest} $\widetilde{H}$ that satisfy (\ref{eqn:21}) and $\abs{V} \leq 2^{\abs{\widetilde{V}}}$, we have:
\begin{align}
\abs{\widetilde{H}} \leq \abs{\widetilde{V}}. \label{eqn:16}
\end{align}
\\

\section{Notation and Definition}
\begin{itemize}
    \item[$\supop{\quad}$:] Super-operators. e.g. $\supop{A}\rho = A\rho A^\dagger$.
    \item[$*$:] See Section~\ref{sect:01}.
    \item[$\zeta$:] See Section~\ref{sect:zeta}.
    \item[$\widetilde{\quad}$:] Generating set. Note that $\widetilde{A}$ means that $A$ can be generated from $\widetilde{A}$, but does not means that $A$ is the \textbf{complete }set of elements that can be generated from $\widetilde{A}$.
    \item[$\expval{\ }$:] The full set of operators that can be generated from the given set. In the context of our paper, the composition rules used in the generation is $*$.
    \item[$G$:] The Pauli operator set.
    \item [$M$:] The noise operator that we want to twirl.
    \item [$V$:] The Pauli basis of the noise $M$.
    \item [$W$:] The twirling gate set. 
    \item [$Q$:] The set of Pauli operators that denote the rows of quotient tables.
    \item [$H$:] The set of Pauli operators that denote the columns of quotient tables.
\end{itemize}

%

\end{document}